\documentclass[reprint,prd,nofootinbib,preprintnumbers]{revtex4-1}
\usepackage{graphicx,psfrag,epsfig,epsf,colordvi,slashed,amsmath}
\newcommand{\beq}{\begin{equation}}
\newcommand{\eeq}{\end{equation}}
\newcommand{\bea}{\begin{eqnarray}}
\newcommand{\eea}{\end{eqnarray}}
\def\be#1\ee{\begin{align}#1\end{align}}

\newcommand{\lsim}{\mbox{\raisebox{-.6ex}{~$\stackrel{<}{\sim}$~}}}
\newcommand{\gsim}{\mbox{\raisebox{-.6ex}{~$\stackrel{>}{\sim}$~}}}

\newcommand{\ov } {\over }

\begin{document}
\preprint{MAN/HEP/2012/016}

\title{Multiple dark matter scenarios from ubiquitous stringy throats}
\author{\bf Diego Chialva$^1$, P. S. Bhupal Dev$^2$ and Anupam
  Mazumdar$^{3,4}$\\} 
\affiliation{
\vspace{1mm}
$^1$Universit\'e de Mons, Service de Mecanique et gravitation, Place
du parc 20, 7000 Mons, Belgium\\ 
$^2$Consortium for Fundamental Physics, School of Physics and Astronomy, University of Manchester,
Manchester, M13 9PL,  United Kingdom\\        
$^3$Consortium for Fundamental Physics, Lancaster University,
Lancaster, LA1 4YB, United Kingdom\\ 
$^4$Niels Bohr Institute, Copenhagen University, Blegdamsvej-17,
DK-2100, Denmark}

\begin{abstract}

We discuss the possibility of having multiple Kaluza-Klein (KK) dark
matter candidates which arise naturally in generic Type-IIB string theory
compactification scenarios.  
These dark matter candidates reside in various throats of the 
Calabi-Yau manifold. In principle, they can come with varied range of masses 
in four-dimensions  
depending upon the hierarchical warping of the throats. 
We show that consistency with cosmological bounds and
four-dimensional effective theory description imposes strong
constraints on the parameter space and the geometry of the throats.
With a rather model-independent approach, 
we find that the mass scales allowed for the KK dark matter particles in various throats 
can vary between 0.1 eV and 10 TeV, depending upon the throat geometry.
Thus, there could be simultaneously more than one kind of
cold (and possibly warm and hot) dark matter components residing in the
Universe. This multiple dark matter scenario could weaken the  
bound on a conventional supersymmetric dark 
matter candidate and could also account for extra relativistic degrees of 
freedom in our Universe.
\end{abstract}

\maketitle

\section{Introduction}

Several astrophysical and cosmological observations have established
the existence of Dark Matter (DM) in our Universe~\cite{silk-review},
and hence, a possible hint for New Physics beyond the Standard Model
(SM). Although all evidences so far are based only on its
gravitational interaction, the combined  
data from cosmological and particle physics sources require any standard DM 
candidate to satisfy the following conditions: (i) it must be stable
on cosmological time scales and have the right pressure (to help large
scale structure formation), 
(ii) it must have the right relic density (not to overclose the 
Universe), and (iii) it must be very weakly interacting
with the SM particles (to satisfy the direct and indirect detection
constraints). 

There exist a plethora of particle  
DM candidates~\cite{feng-review} with masses ranging from $10^{-5}$ eV to as 
high as the Grand Unified Theory (GUT) scale $\sim 10^{16}$ GeV, motivated
by various New Physics scenarios. The most studied DM candidates are
the Weakly Interacting Massive Particles (WIMPs) in the mass range
$\sim$ 1~GeV - 10~TeV, as they occur almost naturally in
very well-motivated particle physics  
theories, and have rich phenomenological implications. For example, one of 
the most popular WIMP candidates is the lightest supersymmetric
particle (LSP) in supersymmetric models with conserved
$R$-parity~\cite{jung-review}  
which were originally proposed to provide an elegant solution to the
gauge hierarchy problem of the SM.  

Various phenomenological extra-dimensional models provide an alternative potential
solution to the gauge hierarchy problem, and also a viable WIMP DM
candidate in the form of the lightest Kaluza-Klein (KK) mode. In the
simplest Universal Extra Dimensional (UED) models, the lightest KK
particle (LKP) remains stable, and hence a possible DM
candidate~\cite{ued-dm}, due to KK-parity which is a remnant of the 
translational invariance in the extra dimension, after orbifolding
(i.e., a $Z_2$ symmetry) is imposed to obtain chirality. 
In the more realistic Randall-Sundrum (RS) model~\cite{RS}, where the hierarchy
between the electroweak and Planck scales arises naturally from a
warped geometry of the extra-dimensional space, 
the stability  of the KK modes can be ensured in a bottom-up approach 
by imposing a gauged symmetry~\cite{agashe}, or in a top-down approach 
by assuming the (approximate) preservation of isometries~\cite{cline}.
 
On the other hand, it is desirable to study a model of DM where the DM candidates 
arise naturally from more complete and fundamental theories capable of also describing,  
in a coherent and unified framework, the other stages of the evolution of our 
Universe, such as inflation~\cite{Dutta} (for reviews, see
e.g.,~\cite{McAllister:2007bg, inflation-rev}). The 
superstring theory is arguably the most developed Planck scale
theory that can make phenomenological connection at low
scales, and hence, can potentially capture the whole history of evolution of 
our Universe. It provides a viable description of our 
four-dimensional spacetime by compactifying six extra dimensions on a
Calabi-Yau manifold. 

In the so-called {\em flux compactification} models, the internal
Calabi-Yau manifold presents background fluxes 
arising from three-forms piercing through the internal cycles, and
numerous sources of energy such as orientifold planes, 
$D$-branes, $\bar{D}$-branes,
$NS$-branes, etc.~\cite{string-review}. The fluxes, branes and
orientifold planes typically reduce the $\mathcal{N}=2$ 
supersymmetry preserved by the Calabi-Yau compactification to
$\mathcal{N}=1$ or null supersymmetry (SUSY). The backreaction from
the fluxes transform the metric from a perfectly factorized one into a
warped geometry. Strongly warped regions develop generically when the
fluxes are supported on cycles localized in small regions of the
internal manifold. A typical example is the case of cycles that can
shrink to a conifold singularity (which can then be deformed), 
arguably the most generic singularity in Calabi-Yau spaces, which can
occur in large numbers due to the presence of many three-cycles.

It is now a prediction of string theory that flux
configurations and  
the generic presence of several conifold points lead to a multiple-throat 
scenario~\cite{march, dimo}.  
Such a setup is also phenomenologically preferred as throats with
different warpings give rise to different hierarchies, thus solving
the hierarchy puzzle in particle physics as well as providing viable
inflationary models in the hidden sectors or in the
bulk~\cite{McAllister:2007bg, Barnaby:2004gg, Chialva:2005zy, tye}. 
However, the string models generically predict many
hidden sectors which makes it difficult to obtain the
branching ratios among visible and hidden sectors after preheating and reheating.

As a consequence of the multi-throat scenario,   
matter localized in the hidden throats
(i.e., different from the throat where the SM resides\footnote{For
  simplicity we will not
consider the case where the SM resides in the bulk.}) 
can represent cosmologically stable DM, as the stability against
decaying into SM fields can be  
ensured by a sufficient suppression of  
decay rates due to the necessary tunneling between throats in a
warped geometry~\footnote{Note that this is different from the
  warped scenarios in~\cite{dimo, cline},  
where the LKP resides only {\it within} the SM throat and its stability is guaranteed
by the (approximate) preservation of isometries of the throat.}~\cite{tye}. 
This, in addition to the usual
visible sector DM candidate in the (supersymmetric) SM throat,
naturally leads to a multiple DM scenario~\footnote{For an incomplete
  list of other multiple DM scenarios in various particle physics
  models, see e.g.,~\cite{otherMDM,DDM}.}. 
We will consider the case of KK modes in the hidden throats from
higher dimensional bulk gravitational fields (which are always present
in the models), assuming for
simplicity no extra branes, and hence, no other matter content in the hidden throats 
although any form of matter in these throats could in principle
represent a DM candidate. The stability and  mass range will depend on
the depth of the throats (gravitational redshifting).

The number of string compactifications is very large (dubbed as the `landscape'). Hence, it would be convenient to find
a general way to check if the multi-throat multiple DM scenario can be effecively
realized. In this paper, we show that the relic
density of the KK modes in sufficiently long throats has remarkably
general features (essentially due to the decoupling of the long
throats from each other), and hence, can be written in a calculable
and rather general way without resorting to the fine details of the
underlying string model. We also study how other quantities of interest
can be analyzed in rather general terms.
In this way, we are able to apply all possible bounds
coming from {\em i}) cosmology/astrophysics and {\em ii}) consistency
of the model description, and derive important 
constraints on the parameter space and the throat geometry even when
the branching ratio of the inflaton to visible and hidden sector
degrees of matter cannot be determined so easily from the string
top-down approach.  
We emphasize again that our phenomenological approach is independent of the details of 
the string inflation models and also of the dynamics of (p)reheating. 

We find that there {\it are} allowed regions of the parameter space
for this multiple DM scenario to work and that some features can be
determined rather generically. In particular, we obtain 
the possible range of masses for the DM candidates residing in various
throats, and the bounds on the local string scales. We also discuss
the bounds on the  
inflationary scale for certain classes of models. Finally, we point out 
some interesting phenomenological consequences of this multiple DM scenario.

 The paper is organized as follows: in Section \ref{stablemodessec}, we discuss
the stability of the KK DM candidates in the multi-throat scenario. In Section
\ref{relic_density_sec}, we calculate their relic density and
obtain the bounds on the parameter space for our multiple DM scenario. 
In section \ref{phenoimpsec}, we briefly discuss some interesting 
phenomenological implications. Our conclusions are given in Section
\ref{conclusion}. 

\section{Stability of the KK Modes in Multiple Throats}\label{stablemodessec} 

In the multi-throat scenario we are entertaining, the new DM
candidates are the KK modes residing in throats {\it different}  
from the SM one~\footnote{We will comment on the possibility of
  having long-lived KK modes also in the SM throat in section
  \ref{relic_density_sec}.  
}.
The stability against decaying to and interacting with the SM particles will be 
ensured by the smallness of the couplings to the SM  
fields, and hence, of the decay rates as they are suppressed by the
necessity of quantum-mechanical tunneling among the separate throats.  
Here we summarize the main features of such tunneling, and in the next section
we will discuss the constraints on the warping factor of the throats  
from {\em i)} cosmological constraints, {\em ii)} consistency of the description
and {\em iii)} the requirements of stability and correct cosmic
abundance of these KK modes.   
 
The local metric for each throat far from the tip can be written  
 as a warped product with the generic form:
 \beq \label{warpedmetrgene}
  ds^2 = H(r)^{-{1/ 2}}(g_{\mu\nu}dx^\mu dx^\nu) +
          H(r)^{{1/2}}(dr^2+r^2 ds^2_{X_5}),
 \eeq
where $\mu,\nu=0,1,2,3$ run through the $4$-dimensional metric.
The radial coordinate $r$ reaches a minimum value of $r_{\text{min}}$ at the
tip of the throat, and the local string scale at the tip of the $i$-th throat 
is given by
 \beq \label{LocalStringScale}
  M_{i} \equiv \left. M_s H(r_{\text{min}})^{-{1}/{4}}\right |_{\text{$i$-th throat}} \equiv M_s h_i,
 \eeq
where $h_i$ is the maximum warping factor of the $i$-th throat
($h_i=1$ corresponds to no warping), and $M_s= \ell_s^{-1}$ is the
10-dimensional string mass scale which is usually taken to be smaller
than the reduced Planck 
mass $M_{\rm PL}= \ell_{\rm PL}^{-1}=2.4\times 10^{18}$ GeV.  

The lightest KK mass will be given by~\cite{dimo} 
 \beq \label{KaluzaKleinMass}
  m_{\text{KK}(i)} \sim \frac{h_i}{R_i}, 
 \eeq
where $R_i\geq \ell_s$ is the curvature radius of the $i$-th throat. 
The spacing between the KK masses is also given by ${h_i \ov R_i}$. 
From the $4$-dimensional
point of view, the largest mass which we can imagine will be always bounded by the
compactification scale,   
while the lowest KK mass can be very small (see e.g.~\cite{Panda:2010uq})
and is limited only by the consistency of the 
model. 

Quantum mechanical tunneling will occur among the separate throats as long as 
the quantum numbers of the tunneling modes match in the
two throats. The decay rate depends on the equation of motion of the
modes, and thus on the specific form of the metric
(\ref{warpedmetrgene}) generating the 
gravitational potential wall between the throats. 

Different modelings of the compactified geometry with different 
form of the warping factor $H(r)$ in Eq.~(\ref{warpedmetrgene}) or
modeling of the bulk region between throats lead to
different tunneling rates. 
The prototypical throat metric is
the Klebanov-Strassler solution \cite{klebanov} for which, however, 
the exact solution of the mode equation is not known. 
Useful approximations, which are adapted to different scenarios, have
been used in the literature in order to obtain the decay rate. 
When the throat can be approximated as an $AdS_5\times X_5$, and the bulk region
connecting two throats as a ``Planck brane'' akin to the case of the
RS model~\cite{RS}, the tunneling rate  
from the $i$-th throat to any other neighboring throat can be estimated to be~\cite{dimo} 
 \beq \label{tunnrateRS} 
  \Gamma_{\text{tunn} (i)}^{\rm (RS)} = (m_{\text{KK} (i)}R_i)^4{r_{\text{min}_{(i)}} \ov R_{i}^2}
  \sim {h_{i}^5 \ov R_{i}}~. 
 \eeq  
Other modelings of the compactified geometry worked out in the
literature lead in general to more suppressed decay rates. For example
those in~\cite{tye} 
present two possibilities giving rise to the decay rates
 \be \label{tunnratesIandII}
  \Gamma_{\text{tunn} (i)}^{\text{(I)}} & \sim h_{i}^9 R_{i}^{-1} 
  \quad \text{and} \quad
  \Gamma_{\text{tunn} (i)}^{\text{(II)}} & \sim h_{i}^{17} R_{i}^{-1}.
 \ee
For concreteness, we will present our results considering these models
as exemplary possible realizations in the  
landscape of compactifications, and investigating what constraints on the 
string models follow from the cosmological and consistency requirements for
a suitable multi-throat DM scenario.

To be a DM candidate the KK modes in the hidden throats must be
stable. Typically models require preserved isometries to guarantee
perfectly stable DM candidates. However, isometries in string throat
compactifications are generically broken because of the gluing of the
throats to the bulk~\cite{Berndsen:2008my}.  
Despite this, we learn from Eq.~(\ref{tunnrateRS}) that tunneling from
a longer throat (small warping $h_i$) to a shorter one is highly disfavored, and
therefore the interactions of the KK-modes in such throat with matter
localized in other throats (as the SM or KK modes in other throats)
will be suppressed (small overlapping of 
wavefunctions). For what
concerns the DM properties, this 
means that the lowest KK mode in the $i$-th throat will be stable against 
interacting with matter localized in other throats provided that the throat is long 
enough to ensure that the tunneling life-time is larger than the age of the 
Universe, i.e., 
\be \Gamma_{\text{tunn} (i)}\lsim H_0
\label{stab}
\ee 
where $H_0=100h~{\rm km}.{\rm s}^{-1}.{\rm Mpc}^{-1}$, and the parameter $h$ 
has been measured by WMAP to be $h=0.72\pm 0.03$~\cite{wmap}.  

The KK modes in any given throat could still be unstable due to (i)
self-interactions, i.e., interactions with matter localized in the
same throat, and (ii) annihilation into massless bulk
modes (gravitons)\footnote{The decay $1 KK_{(i)} \to 2
  \text{gravitons}$ is forbidden by Kaluza-Klein number conservation
  for our gravitational KK non-zero modes. The annihilation $2 KK_{(i)} \to 1
  \text{graviton}$ turns out then to be the dominant interaction left to consider,
  as we have already discussed the tunneling effects that suppress
  the interactions involving modes from different throats or the SM.}. 
For what concerns the interactions (i), as already noted in the
introduction, for a simple viable multiple DM 
scenario, we have assumed the hidden throats not to harbor any matter
content other than the KK modes. The latter self-interact, with
a suppression scale given essentially by the local string
scale~\cite{Chialva:2005zy, tye} because of their localization, until they 
freeze-out and eventually become DM relics. In section
\ref{relic_density_sec}, we will see that, in the
absence of any  
isometries making them absolutely stable,
this process is very rapid for the values of parameters
compatible with suitable DM scenarios and that essentially only lowest
KK modes remains in the hidden throats as DM relics. 
In the case of the visible throat (containing the SM degrees of 
freedom), the localized KK modes should rapidly decay into the SM 
radiation to provide a standard cosmology with radiation domination at
early times. This will also be discussed in Section~\ref{relic_density_sec}.

On the other hand, the
annihilation into massless bulk modes (ii) is suppressed because of two
reasons~\cite{Chialva:2005zy, tye}:  
first, the annihilation cross-section of the KK modes to gravitons is
Planck-suppressed as
$\sigma_g \sim M_{\text{PL}}^{-2}$ (whereas the KK self-interactions
are warp-enhanced, and thus essentially suppressed by the local string
scale $\sigma_{KK} \sim g_s^2
M_{i}^{-2}$, and further reduced by the 
fact that only modes carrying oppositely conserved internal momentum can
annihilate. Second, gravitons are rapidly redshifting ($\sim a^{-4}$, $a$ being the scale factor)
compared to the massive KK modes ($\sim a^{-3}$), when the
latter become non-relativistic. Therefore, the gravitational interactions 
become ineffective soon. This also ensures that gravitons are not 
overproduced~\cite{tye}.

Thus we conclude that in the
multi-throat scenario considered here, ensuring the condition
(\ref{stab}) is sufficient for the stability of the KK modes in a
given throat. Ensuring that condition will put constraints on the
geometry of the compactifications.    

\section{Bounds and relic density of the KK Dark Matter}\label{relic_density_sec}

The production of the KK modes localized in various throats
depends on the stringy realization of inflation and
(p)reheating. 
However, the post-production evolution of the KK modes in
the  hidden throats shows remarkable
model-independent features, because of the particular background 
geometry leading to the tunneling-suppression of the interaction among different
throats. 

This scenario has a striking difference compared to the
standard WIMP DM picture.  
In the latter case, the DM candidate participates in the {\em same}
thermal history as the 
SM degrees of freedom and evolves
independently only after freeze-out (which
determines its relic density). 
Instead, in the multi-throat scenario, the DM particles in different
long enough throats
are decoupled from each other and from the SM degrees of
freedom by tunneling suppression of the interaction  
cross-sections between them~\footnote{Gravitational
  interactions, though not tunneling suppressed, are 
suppressed by $M_{\rm PL}^2$ and hence unable to lead to thermal
  equilibrium among the throats.}. 
As shown later in this section, their relic density is 
determined essentially by the initial conditions~\footnote{In
  the multi-throat scenario, the DM 
in each throat has its own thermal history, and in
case it reaches thermal equilibrium in the throat, its own
temperature. These possible temperature differences might affect the
counting of the total relativistic degrees of freedom $g_*$ and $g_{*S}$ 
entering respectively the formulas for the energy density and entropy. However, the
dependence of the relic density on these quantities will turn out to be
weak.}. 

Such dependence on the initial conditions of production can be
efficiently parametrized by subdividing the
models into two general classes, as discussed below, on the basis of
the mechanism of reheating and production of the KK modes. 
Therefore, our analysis can be rather model-independent and adapt for
the landscape setting.

In one class of models, the energy which excites the KK modes in the hidden
throats and reheats the SM in its own throat is initially localized
essentially in a single (non-SM) throat, from which it is transferred to the
others by {\em tunneling}. A typical example is the  $D$-$\bar D$
inflation models with the mechanism of reheating-through-tunneling out
of the inflationary throat
studied in \cite{Chialva:2005zy, tye}
. Note however that our analysis is 
independent of the specific models, as appropriate for the landscape:
it applies to all models where the 
energy responsible for reheating of the DM and SM fields resides
initially in a {\it single} throat, independently of the details of
how and where inflation occurred.
We will call this throat the ``initial throat'' (indicated with a
label ``in''), and
this class of models ``throat-reheating''
~\footnote{We will not
  consider the case when the SM and the 
initial throat coincide, as the analysis in this case is
similar to that in~\cite{agashe,cline} for the RS-model.
}.

In the other class of models, which we call ``bulk-reheating'', the 
energy responsible for reheating does not reside in
one specific initial throat, but is spread 
over the internal manifold. Typical examples are those where the
reheating field is a bulk inflaton or a bulk modulus, which decays
releasing its energy all over the 
warped Calabi-Yau manifold. These models, however, must reheat
the SM efficiently, therefore requiring different branching
ratios among the throats, and also among the bulk and the throats. Note again 
that our analysis will be independent of specific models in this
case too. It may even be 
the case that inflation has occurred in one specific throat, but
couples more efficiently to bulk moduli or fields such as gravitinos,
which then, through their decay, 
will reheat the other degrees of freedom from the
bulk.

As our analysis will be independent of the specific
features of perturbative reheating or preheating, we
use the term ``reheating'' in a generic way, encompassing all possible
ways that produce the KK modes within the hidden throats.

Summarizing, the essential features of the models relevant for our analysis
are: 1) sufficiently long throats are decoupled from each other because of
tunneling suppression, and 2) the models can be divided in two broad
classes depending on if 2A) the energy responsible for reheating is
initially localized in one specific throat from where it has to tunnel
out, or if 2B) the energy responsible for reheating can be released
through direct couplings and does not require tunneling to
excite the other degrees of freedom.

\subsection{Reheating in one throat}\label{3a}

In this scenario the energy density responsible for reheating is
initially localized in one specific throat from where it has to tunnel
out. The typical example is the scenario of $D$-$\bar D$ inflation in one throat.
 In that case, the end of inflation is brought about by the
 annihilation of the $D$ and $\bar D$ branes. Their
decay products are massive fundamental closed strings, which decay
until KK modes that quickly thermalize are left
over~\cite{Chialva:2005zy, tye}. Then these modes 
tunnel to the other throats 
reheating them~\footnote{\label{tunnnonrelmodes}The
modes at tunneling are non-relativistic. Indeed, if we assumed
that they were relativistic, the corresponding 
temperature $T_{\rm in,tunn}$ would be found to be smaller than the mass of 
even the lightest KK mode ($m_{KK(\rm in)}$). For example, even for
the less suppressed decay rate in the case of RS-like throat geometry,
the ratio is 
${T_{\rm in,tunn} \ov m_{KK(\text{in})}} =
\sqrt{R_{\text{in}}M_{\text{PL}}}\,h_{\text{in}}^{{3/ 2}}$, which is
generally much smaller than one.}.
 
The largest part of the energy  will be tunneled into the longest
throat since it has the densest spectrum, which favors tunneling as the level 
matching is easier and more energy levels fall within the width of the
initial tunneling mode. The remaining degrees of freedom 
end up in sufficiently long throats since the tunneling
out of them is very suppressed. The shallow throats, having a less dense 
spectrum, receive less likely the energy tunneling from the
initial throat, and moreover the received energy
will anyway tunnel out of them in a short time-scale due to their large 
tunneling rates. Hence, the shallow throats can be practically treated as 
empty for the purposes of our analysis.    

Note that since the initial throat has to be shallower than
the other throats for tunneling to occur, even the lightest modes in the
initial throat will tunnel to modes which are not the
lightest ones in their respective throats because of the
level-matching.  
These higher modes in the final ($i$-th) throat are in general unstable and
decay into the lighter 
ones much more rapidly than the Hubble rate after tunneling from the
initial (in) throat provided that the self-interaction rate of the KK
modes in the $i$-th throat 
$\Gamma_{\text{SM}} = {m_{\text{i}}^3 \ov M_{i}^2}$ is
larger than the Hubble rate right after 
tunneling\footnote{Due to level matching, the mass of
  the tunneled KK modes is $m_{\text{i}} \sim m_{\text{KK(in)}}$.}, i.e.,   
 \beq
   h_{\rm in}^3 >  R_{\rm in}^2 M_i^2 
    \sqrt{R_{\rm in}^2 M_{\rm PL} \Gamma_{\text{in, tunn}}}
 \eeq
which, as we will see later, is easily satisfied for the
parameter space leading to allowed DM models. 
Hence, we can safely assume that the long hidden
throats harbor only DM candidates in the
form of the lightest KK modes.

Assuming that the SM radiation dominates soon after tunneling
and is in thermal equilibrium (the necessary
conditions will be discussed later), as envisaged in a standard cosmology with 
radiation domination at early times followed by matter domination, and
recalling that the different long enough throats are decoupled from 
each other by tunneling suppression,
we can derive an equation for the current relic density of the KK dark matter
candidates in the {\em hidden throats}:
 \bea \label{abun-open}
  \Omega_X h^2 
   & = & {1.3 \times10^{-4} \ov g^\prime_*\, g_{*}(T_{\text{in, tunn}})}
  {\sum_i m_{X_i} n_{X_i}(t_{\text{in, tunn}}) \ov T_{\text{in, tunn}}^3 T_0}
   \nonumber \\
   & = &   
   {1.8 \times 10^{8} \ov g^\prime_*} \biggl(\frac{T_{\rm in,tunn}}{1~{\rm GeV}}\biggr)
   F_X
 \,.
 \eea
Here, $T_0$ is today's temperature, $t_{\rm in,tunn}/T_{\rm in,tunn}$ is the
time/temperature after tunneling from the initial throat,  
$g^\prime_* \equiv {g_*(T_0)g_{*S}(T_{\text{in, tunn}}) \ov g_{*S}(T_0)g_{*}(T_{\text{in, tunn}})}$  
(with $g_*, g_{*S}$ the total relativistic degrees of freedom
entering respectively the relation for energy density and entropy),
$n_i(t)$ is the number density of the DM particles $X_i$, 
and  
 \beq \label{FX}
 F_X \equiv \sum_i\left(m_{X_i} \ov m_{KK({\rm in})}\right)
  \left(n_{X_i}(t_{\text{in, tunn}}) \ov n_{\text{tot}}(T_{\text{in, tunn}})\right)
 \eeq
is the sum of the branching ratios after tunneling from the
initial throat for the $X_i$ modes
in the throats $i \neq \{{\rm SM, in}\}$ weighted
by the ratio of their masses $m_{X_i}$ and the lightest KK mass
$m_{KK({\rm in})}$ in the initial throat\footnote{Note that in
  deriving Eq.~(\ref{abun-open}) 
  we do not make
  any assumption (relativistic/non-relativistic or even thermal/non-thermal) on the
  energy or number density of the modes $X_i$ at $t_{\text{in, tunn}}$. 
   The final expression of the relic density in Eq.~(\ref{abun-open}) is however 
similar to those in~\cite{Campos:2001zb, Allahverdi:2002nb}.}. 
Note that the sum in Eq.~(\ref{FX}) includes only the non-SM long
throats; the dynamics of Kaluza-Klein modes in the SM throat requires
a different analysis that take into account the interactions within
the SM fields, as shown later in this section. 

\begin{figure}[b]
\centering
\includegraphics[width=6cm]{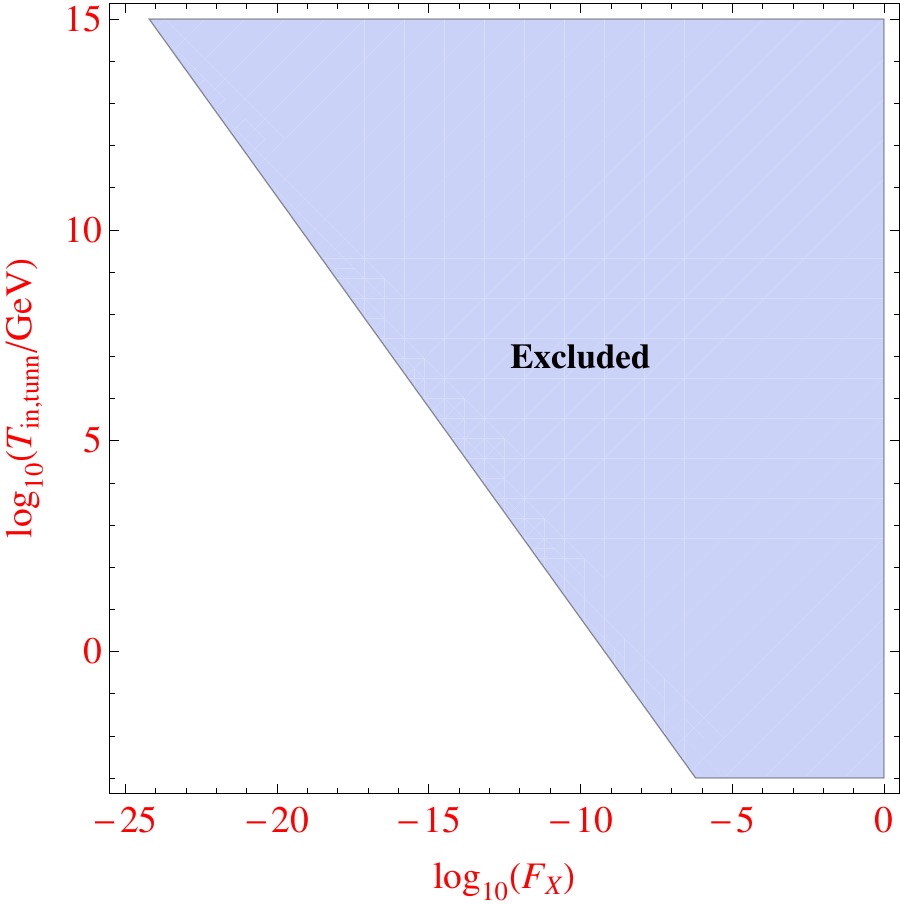}
\caption{The allowed parameter space satisfying the overclosure
  constraint. We have taken $g_{*S}(T_{\text{in, tunn}})\simeq
  g_{*}(T_{\text{in, tunn}}) = 250$.} 
\label{fig:1}
\end{figure}

In order to avoid overclosure of the Universe by the KK dark matter particles, 
 we require the relic density given by Eq.~(\ref{abun-open}) to be less than
the $3\sigma$ upper limit of the WMAP  
observed value~\cite{wmap}: 
$\Omega_Xh^2\leq 0.13$. The allowed range of $F_X$ satisfying this
condition is shown in Fig.~\ref{fig:1}.  This result will be used
later in this section to constrain the parameter space for a multiple
DM scenario. As the dependence on the total 
numbers of relativistic
degrees of freedom at different temperatures
is not strong, we have assumed the common value 250 at early times,
to account for the Minimal Supersymmetric SM (MSSM) sector plus some hidden degrees of freedom.  

The condition of SM radiation domination after tunneling from the
initial throat requires the
SM throat to be the one which receives more energy from the initial
tunneling. A necessary 
condition for this is that the SM throat must have the densest spectrum,
which favors tunneling as the matching between
energy levels is easier (tunneling is favoured when the final energy
eigenstates fall within the width of the initial state). This implies
 \beq \label{warpsmallcond}
   h_{i}R_{i}^{-1} > h_{\text{SM}}R_{\text{SM}}^{-1}.
 \eeq
The radiation domination further depends on the
thermal history of the degrees of freedom in the SM throat. The modes  
tunneled there
 can 1) decay into the lowest KK(SM) modes, 2) thermalize, 3)
decay into SM fields.  In the allowed 
range of warping, the decay into the 
SM radiation~\cite{Chialva:2005zy, KKtoSM} is very rapid, i.e., less
than one Hubble time  
$\sim H^{-1}_{\text{in, tunn}}
    \sim \Gamma_{\text{in, tunn}}^{-1}$ right after they have tunneled, 
provided that
   \beq \label{SMdecay}
     \Gamma_{\text{SM}} = {m_{\text{KK, SM}}^3 \ov M_{\text{SM}}^2} 
      = h_{\text{SM}}{\ell_S^2 \ov R_{\text{SM}}^3}
      \gg \Gamma_{\text{in, tunn}}.  
    \eeq
Because of Eq.~(\ref{warpsmallcond}),
this also entails a lower bound on the warping of the hidden
throats. 
Note that the breaking of isometries due to the gluing of the throat
to the bulk~\cite{Berndsen:2008my} ensures the decay of the Kaluza-Klein modes in the
Standard Model throat.

It is possible that mildly broken isometries could exist
at the tip of the initial throat~\cite{Kofman:2005yz}. In this case, 
the KK(in) modes possessing the
related approximately conserved quantum number would
be long-lived (LL) modes. They would not easily tunnel out
because of difficult quantum number matching between throats, so that
the other DM and SM throats will not be populated by these dangerous
LL modes. Their present relic density 
reads
 \beq \label{relicLLinf}
  \Omega_{\rm LL_{\text{in}}} h^2 = {3.24 \times 10^{26} \ov g_s^2 \widetilde g_{*} \bar g_{*}}
    {M_{\text{in}}^2 \ov M_{\rm PL}^2}
   {T_{\rm in, tunn} \ov m_{X_{\rm in}}}
   \biggl(\frac{m_{\rm LL_{in}}}{T_{\rm dec, LL_{in}}}\biggr)^{\!\!{3 \ov 2}},
 \eeq
where $T_{\rm dec, LL_{in}}$ is the decoupling temperature of the LL
modes with mass $m_{\rm LL_{in}}$, given by
 \beq \label{TdecLLinf}
  \frac{m_{\rm LL_{in}}}{T_{\rm dec, LL_{in}}} = \log\biggl({\sqrt{3} \ov (2\pi)^{{3 / 2}}}g_{*}(T_{\rm dec, LL_{in}})^{{1 \ov 2}} g_s^2\frac{M_{\rm PL} m_{\rm LL_{in}}}{M_{\rm in}^2}\biggr),
 \eeq
and $M_{\text{in}}$ is the local string scale in the
initial throat. We have also defined
 \be 
  \widetilde g_{*} & \equiv 
   {g_{*S}(T_{\rm dec, LL_{in}})g_*(T_{\text{RDMD}})^{{3 \ov 4}} \ov g_{*}(T_{\rm dec, LL_{in}})^{{1 \ov 2}}g_{*S}(T_{\text{RDMD}})}
  \\
  \bar g_{*} & \equiv {g_{*}(T_{\rm in, MD})g_{*S}(T_{\text{in, tunn}}) \ov g_{*S}(T_{\rm in, MD})g_*(T_{\rm in, tunn})},
 \ee
where $T_{\rm in, MD}$ is the temperature when the lightest KK mode
in the initial throat becomes non-relativistic, and
$T_{\text{RDMD}}$ is
the temperature at the standard transition between radiation and
matter domination.
Eqs.~(\ref{relicLLinf}) and (\ref{TdecLLinf}) match the standard
result for cold relics~\cite{KolbTurner}, except for the 
additional suppression factor 
$\text{{\small $(\bar g_*)^{-1}$}}{T_{\text{in, tunn}} \ov
  m_{X_{\text{in}}}}$ due to the matter domination (MD) period between
the time when the lightest KK modes in the initial throat become non-relativistic
and the end of tunneling from the initial throat.

\begin{figure}[h!]
\centering
\includegraphics[width=8cm]{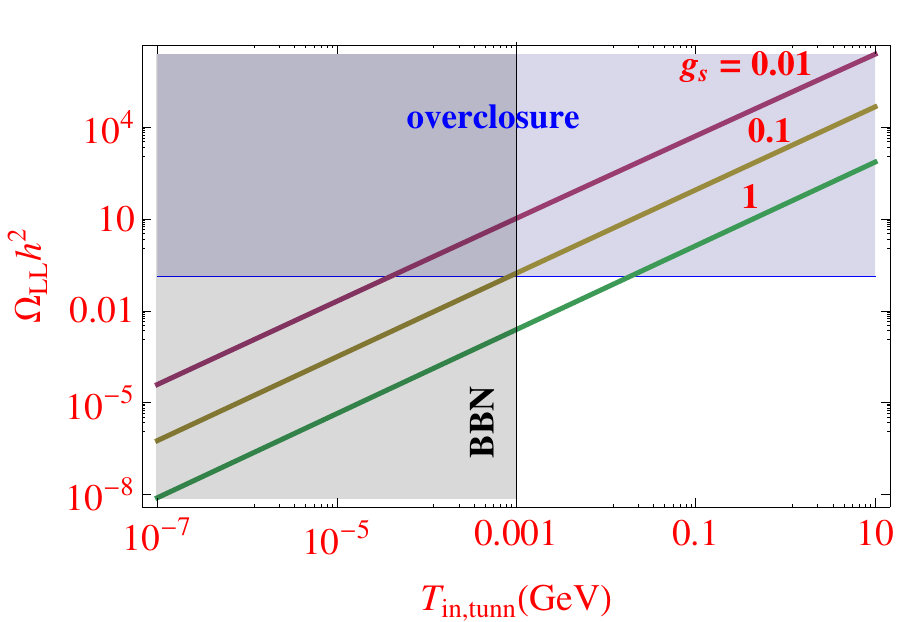}
\caption{The relic density of the long-lived KK modes in the
  initial throat with RS-like geometry as a
  function of the tunneling temperature. Here we have chosen 
  $R_{{\rm in}} = 10 \ell_s$ and $M_s = 10^{-2}M_{\rm PL}$.  The 
  vertical shaded region is excluded from BBN constraints. In the
  horizontal shaded region, these LL modes will overclose the
  Universe. Note that for the string coupling $g_s\leq 0.1$,
  these LL modes, if present, are always dangerous relics.} 
\label{fig:2}
\end{figure}

To illustrate our point on the LL modes, the relic density
(\ref{relicLLinf}) for the LL(in) modes is plotted in Fig.~\ref{fig:2}
for some typical   
string coupling values. In this plot, we show the case where the throat geometry is
well-approximated by the $AdS_5 \times X_5$ RS-like picture, and we have used 
$R_{{\rm in}} = 10 \ell_s$ and $M_s \sim 10^{16}$ GeV (GUT
scale). 
As before, due to the weak dependence on the
total numbers of relativistic
degrees of freedom at different temperatures,
we have assumed the common value 250 at early times.

We see that unless the interactions of
these LL modes are strong enough (for $g_s\gsim 1$),  
their relic density will be too large in the allowed temperature
range, i.e. $T_{\rm in, tunn}, T_{\rm dec, LL_{in}}$ larger than the Big
Bang Nucleosynthesis (BBN)
temperature $\sim 1$ MeV~\cite{KolbTurner}. Thus, we conclude that for
reasonably small values of 
the string coupling $g_s\leq 0.1$, we must not have the long-lived KK
modes in the initial throat. In other words, all the
isometries must be broken (which is anyway expected by the junction of
the throats to the bulk~\cite{Berndsen:2008my}) 
in order for the Universe not to be overclosed. Increasing the value of
$R_{\text{in}}$ worsens the situation, and only a lower value of $M_s$
would improve it. For the cases where the 
decay rate is more suppressed, as in the models of \cite{tye},
the constraint for the LL modes is even more difficult to satisfy; for
instance, for the values of the 
parameters chosen for Fig.~\ref{fig:2}, even the case $g_s = 1$ is
ruled out for the LL modes. 

Beside overclosure, there are other cosmological constraints 
to be satisfied as well.
First of all, the tunneling decay must occur well before the
BBN in the SM throat which happens at temperature  
$T_{\rm BBN} \sim 1$ MeV, and sets the lower bound for 
$T_{\rm in,tunn}$. We should mention here that this is the only robust 
lower bound we can put on the tunneling (and hence, reheating)
temperature since we do not yet know for sure the exact history of the
observable Universe before the epoch of BBN. In practice, such extreme
values for reheating temperature could be problematic 
for successful baryogenesis (see however~\cite{Davidson:2000dw} for an
example where this could still work). In such cases, the allowed
parameter space to be  
shown in the following Figures might shrink considerably.  

Furthermore, for our four-dimensional field theory description to be
valid, we should have the temperature $T_{\text{in, tunn}}$
lower than the local string scale in each throat in order to avoid exciting
stringy degrees of 
freedom, which entails a lower bound on the local string scale (or, on
the warp factor, and thus, on the KK mass) in each throat: 
 \beq \label{open-h}
  M_i \gsim T_{\text{in, tunn}} \Rightarrow 
   m_{X_i} \gsim {\ell_s \ov R_i}T_{\text{in, tunn}}.
 \eeq
It may be noted here that the more realistic lower bound on the warp
factor might be obtained by considering the scale of 
throat deformation/decompactification due to moduli
destabilization. However, this is   
very hard to compute explicitly in the Hubble expansion background
~\cite{maharana,Frey}, and hence, we use the
more intuitive (and in general weaker) bound in Eq.~(\ref{open-h})
using the string scale. 

{\em Summarizing}, 
the bound from BBN temperature, the condition Eq.~(\ref{open-h}),
the  relic density overclosure constraint, the requirements of
rapid and efficient SM reheating (i.e., 
Eqs.~(\ref{warpsmallcond})-(\ref{SMdecay})
), and the stability condition (\ref{stab}) in each throat
are the conditions that need to be satisfied in any working multi-throat DM
model. These in turn 
put strong constraints on the warping factors of the throats
harboring the KK modes, which
are directly related to the DM masses $m_{X_i}$ and the local string
scales $M_i$ in the throats. 

We have studied the allowed region of parameter space
after imposing these conditions. The stringy parameters in the
landscape setting are warpings,
throat curvatures, scales, tunnelings rates, and initial branching
ratios. Through the above equations and discussion, we have related them 
in a rather model independent way (that is, independently of the fine
details of the compactification) to the constrained quantities such
as relic densities and KK masses. The
relation between the warping $h_{\text{in}}$ and the tunneling temperature 
$T_{\text{in, tunn}}$ follows 
from the dependence of $\Gamma_{\text{in, tunn}}$ on
$h_{\text{in}}$, which in turns depends on the specific form of the throat
metric. As mentioned before, we will consider the decay rates in Eqs. 
(\ref{tunnrateRS}) and (\ref{tunnratesIandII}) as exemplary model 
realizations.

\begin{figure*}[ht!]
\centering
\includegraphics[width=5.4cm]{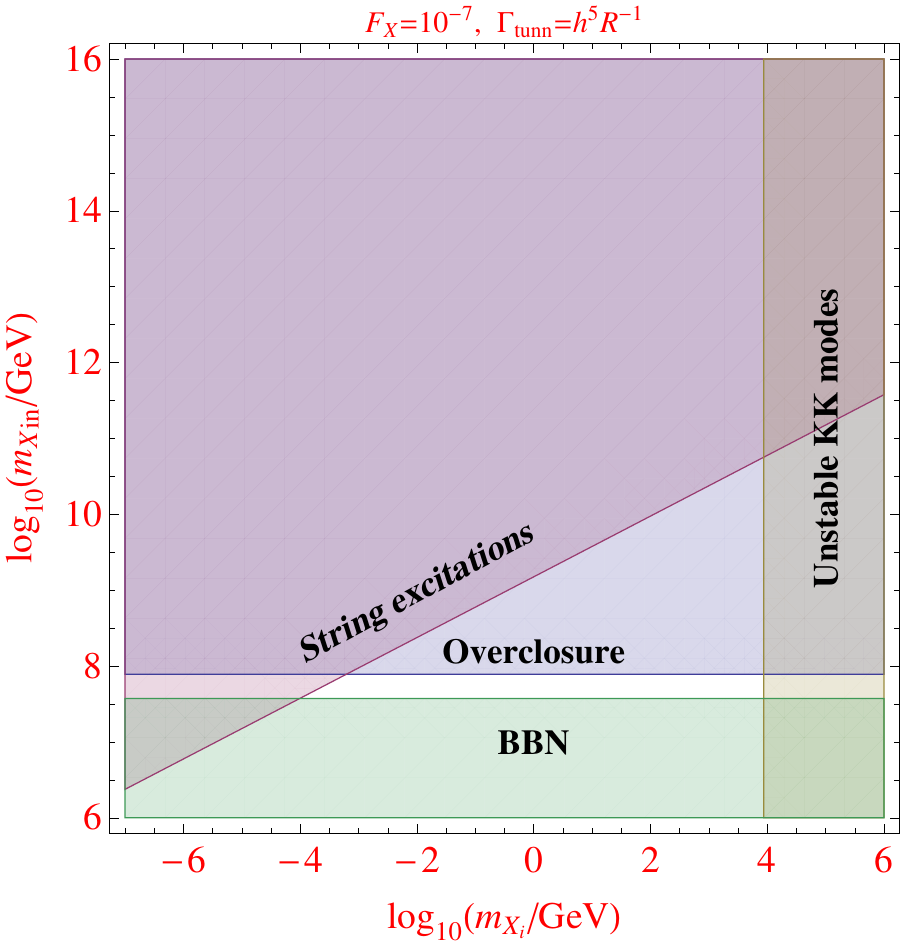}
\includegraphics[width=5.4cm]{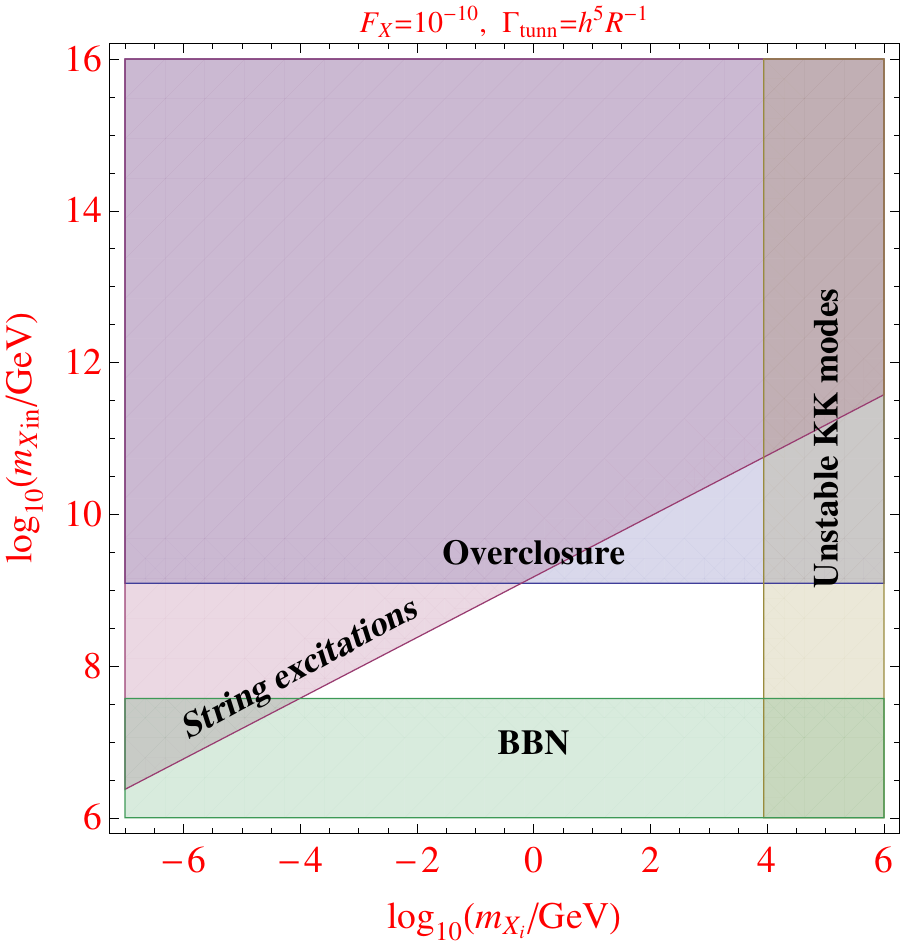}
\includegraphics[width=5.4cm]{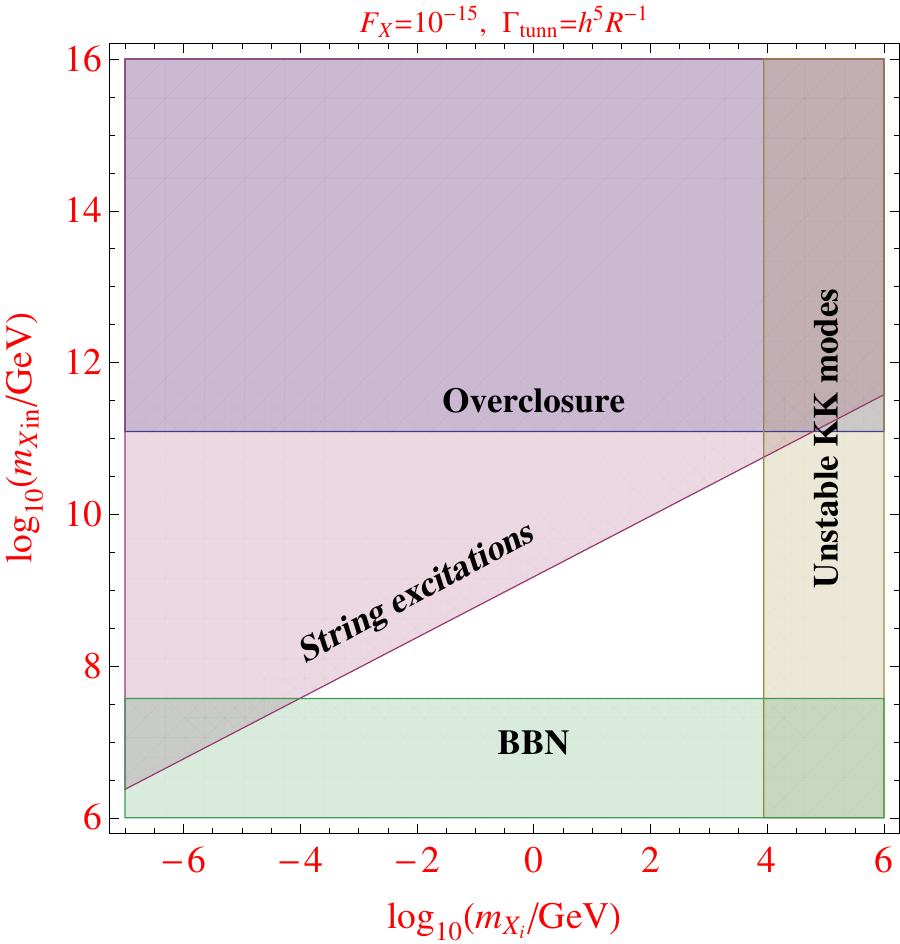}
\caption{Examples of allowed ranges of parameters, shown as the white
  (unshaded) region, for the class of models discussed in section \ref{3a}
  for different fractions 
  $F_X$ defined by Eq.~(\ref{FX}) and arbitrary number of throats in
  terms of the lightest KK masses in the initial throat ($m_{X_{\rm in}}$)
  and other hidden throats ($m_{X_i}$). We
  plot here the case of $AdS_5 \times X_5$ RS-like throat geometry (for other cases see text).  
  Here, we have chosen the string scale to be $M_{\rm PL}/100$ and the $AdS$
  radius of curvature in each throat to be $10M_{\rm PL}^{-1}$, where
  $M_{\rm PL}=2.4\times 10^{18}$ GeV is the reduced Planck mass. Note
  that the condition (\ref{warpsmallcond}) in combination with
  (\ref{SMdecay}) is automatically satisfied {\em for these
  examples} and we have not shown it in the plot.} 
\label{fig:3}
\end{figure*}

For illustration, we present in  
Fig.~\ref{fig:3} the results, for different fractions $F_X$ and certain
values of the string and throat curvature scales,
for the case when the throat geometry is
well approximated by a RS-like $AdS_5 \times X_5$ picture, and will discuss
later the results for other parameter values and models with more
suppressed tunnelings as those 
in \cite{tye}. 

In Fig.~\ref{fig:3} the constraints are shown as various shaded regions and 
the allowed parameter space after imposing all the constraints is
shown as the white (unshaded) region. We see that the size of the
allowed region (which gives a measure of the fine-tuning) depends on
the value of $F_X$ in the landscape of compactifications, but some
features of the working models (such as the DM mass range) do not.
We have used $R_{i} \simeq R_{{\rm in}} \simeq R = 10 \ell_s$ and $M_s
\sim 10^{16}$ GeV (GUT scale)~\footnote{\label{equalradii} The 
assumtion that the curvature radius $R$ is similar for 
all throats is justified by the fact that in string theory flux
compactification typically
$R \sim (F)^{{1 \ov 4}}g_s\!^{{1 \ov 4}} \ell_s$ \cite{string-review,
  klebanov, gidd} (this is certainly true 
when the throat metric is of the $AdS_5 \times X_5$ RS-like kind),
where $F$ is given by some flux/brane 
numbers which can vary from throat to throat. Even if $F$
varies by a factor $10^6$, $R$ differs just by about one order of
magnitude.}. With the above choices, the allowed mass range for the DM
candidates is limited to about $0.1$ MeV - $10$ TeV, with the local
string scale in the initial throat at most at $\sim 10^{12}$ GeV. Note
that the condition (\ref{warpsmallcond}) in combination with
(\ref{SMdecay}) is automatically satisfied for the cases shown in
Fig.~\ref{fig:3} and we have not shown it in the plot. However,
this constraint is generally important in the other cases, discussed below. 

In the case of models where the tunneling rate is more
suppressed by higher powers of the
warping than in the case of RS-like geometry of the throats 
(see section~\ref{stablemodessec}), the
allowed region of parameter 
space shrinks, and the warping of
the initial throat has to be 
shallower, i.e., $m_{X_\text{in}}$ and $M_{\text{in}}$ have to be
larger. For the most suppressed case with $\Gamma \sim h^{17}
R^{-1}$ for the throats, the mass scales are found to be quite
restricted; for instance, $m_{X_\text{in}}$ can only vary within about
an order of magnitude. 

As mentioned, we have also studied how the allowed region of parameter space
varies with the parameters $R$ and $M_s$. Keeping the latter fixed and
letting $R$ vary, the allowed region shrinks (both for RS-like throat
geometry and for the cases with more suppressed decay rates) and moves to
lower mass scales 
However, we find that the mass of the DM candidates cannot
be lower than about $0.1$ eV satisfying all the constraints along with the condition in 
Eq.~(\ref{SMdecay}). 
Also, keeping $R$ fixed, the upper bounds on the KK mass scales in the
initial and other hidden throats get lower
when $M_s$ gets smaller (without affecting much the lower bound on $m_{X_i}$), and
the allowed region again shrinks. 

Interestingly, when the throat geometry is 
well approximated by an $AdS_5 \times X_5$ RS-like picture, for all
values of $R$ and $M_s$, we find
that no allowed region exists for the  
local string scale in the initial throat $M_{\text{in}}$ exceeding about $10^{13}$ GeV
in the compactified effective theory (i.e., with stabilized throat
geometry). If inflation take places in this throat, then, its scale cannot be
larger than the local string one, and this would rule out a high scale
(around GUT-scale) multi-throat inflationary scenario in this
case. This in turn puts a strong constraint on the tensor-to-scalar ratio, and hence, 
the possibility of detecting primordial gravitational waves (see
e.g.~\cite{inflation-rev}).
 
\subsection{Reheating in the bulk}\label{bulkreheatingsec}

In this scenario the energy density responsible for
reheating is not localized in a throat and couples 
without tunneling suppression to all throats. Although this initial
energy density can be stored in bulk modes or moduli fields different
than the inflaton, for simplicity we will call ``inflatons'' all the
degrees of freedom associated with the initial energy density.
The reheating process in this kind of models can lead to various
effects, including destabilization of the geometry. For our purposes,
however, we can start our analysis from when the throats have again
settled down and the perturbative Kaluza-Klein picture is valid. 
We assume that the KK modes have been produced by the inflaton
reheating process at a reheat temperature $T_{R}$ within the throats
where they are localized.  

As discussed earlier, the KK modes in different
long throats will be
decoupled from 
each other 
because of tunneling suppression of the relevant
interaction cross-sections.  
The present DM abundance can be estimated, assuming SM radiation
domination after reheating, and without making assumptions regarding
the energy/number density of the $X_i$ modes at the reheating time
$t_{R}$. We obtain a result similar to
Eq.~(\ref{abun-open}),   
with $T_{\rm in, tunn}$ replaced by $T_R$ and 
$F_X = {\sum_i m_{X_i} n_{X_i}(t_{\text{R}}) \ov \rho_{\text{tot}}(T_{\text{R}})}$, 
where $\rho_{\text{tot}}(T_{\text{R}}) = 0.3g_*(T_R) T_R^4$. 

Differently than the throat-reheating scenario, in the bulk-reheating
case we are discussing here, the SM fields can be reheated directly since the 
``inflaton'' is directly coupled to them. 
In this case, we do not have to require the condition 
(\ref{SMdecay}) for SM radiation to be produced rapidly. 
However, the inflaton is coupled to the other throats as well and with
the bulk; hence, there could easily be overproduction of the non-SM
degrees of freedom and/or gravitons.   
The efficiency of these processes is however model-dependent and
 beyond the scope of our discussion here~\footnote{For instance, in
  large volume compactification models, it is well-known that the
  inflaton branching ratio to the hidden degrees are much more than
  the visible sector degrees of freedom~\cite{Cicoli}.}. 

Another important difference with the throat-reheating case 
concerns the possibility of long-lived modes 
because of mildly broken isometries~\cite{Kofman:2005yz}.
We recall  that in the throat-reheating case all the energy density
comes from the initial throat by tunneling, and LL modes
in that throat will not easily tunnel out
because of difficult quantum number matching. Thus, the DM and
SM throat will not be populated by dangerous long-lived KK modes. 
But in the case of bulk reheating, dangerous LL
modes can be excited in all throats, if mildly broken isometries
exist, due to the direct coupling with the reheating (inflaton)
sector. 
Their relic density in the SM throat has the same form as 
Eq. (\ref{relicLLinf}) {\em without} the suppression factor due to
matter domination before tunneling,
and substituting all  ``in"-labels with the label ``SM'' indicating the
throat in question: 
 \be \label{relicLL}
 \Omega_{\rm LL_{\text{SM}}} h^2 = 
 {3.24 \times 10^{26} \ov g_s^2 \widetilde g_{*}}
    {M_{\text{SM}}^2 \ov M_{\rm PL}^2}
   \biggl(\frac{m_{\rm LL_{SM}}}{T_{\rm dec, LL_{SM}}}\biggr)^{\!\!{3 \ov 2}},
 \ee
where the decoupling temperature is obtained by similarly substituting the 
``in"-labels with ``SM'' in Eq.~(\ref{TdecLLinf}). From Eq.~(\ref{relicLL}), 
we can see that in the presence of LL modes in the SM throat, the overclosure
bound will constrain the warping factor $h_{\rm SM}$. For instance, 
for $M_s \sim 10^{16}$ GeV, $g_s \sim 0.1$ and $R_{\rm SM}=10\ell_s$,
we get $h_{\rm SM}\lsim 10^{-14}$. The bound gets stronger if the
interactions of the LL modes are weaker (i.e., $g_s$ is smaller or $M_s$ is increased).
For each of the other hidden throats, the LL(j) relic density will be determined by the
effective decoupling among throats due to tunneling suppression.

We have therefore two scenarios for
the requirement that stable KK modes do not overclose the Universe in
the bulk reheating case, 
depending on if the mildly broken
isometries responsible for the presence of long-lived modes exist or
not. In the latter case, the total relic density is just $\Omega_{\rm X}h^2$ given by
Eq.~(\ref{abun-open}) with $t_{\rm in, tunn}/T_{\rm in, tunn}$ replaced by $t_R/T_R$.
But if there are some isometries and the relative LL modes, the total
relic density is obtained by including in 
$F_{\rm X}$ the number density and mass of 
the LL(j)-modes. 

As before, we ask here for the temperature $T_R$ to be lower than the
string scale in the throats, to avoid exciting stringy degrees of
freedom, which entails $M_i\gsim T_R$. 
This, together with the relic density overclosure constraint and the
requirement of stability of the KK modes, leads to bounds on the mass
scale of the KK modes.  

\begin{figure*}[ht!]
\centering
\includegraphics[width=5.4cm]{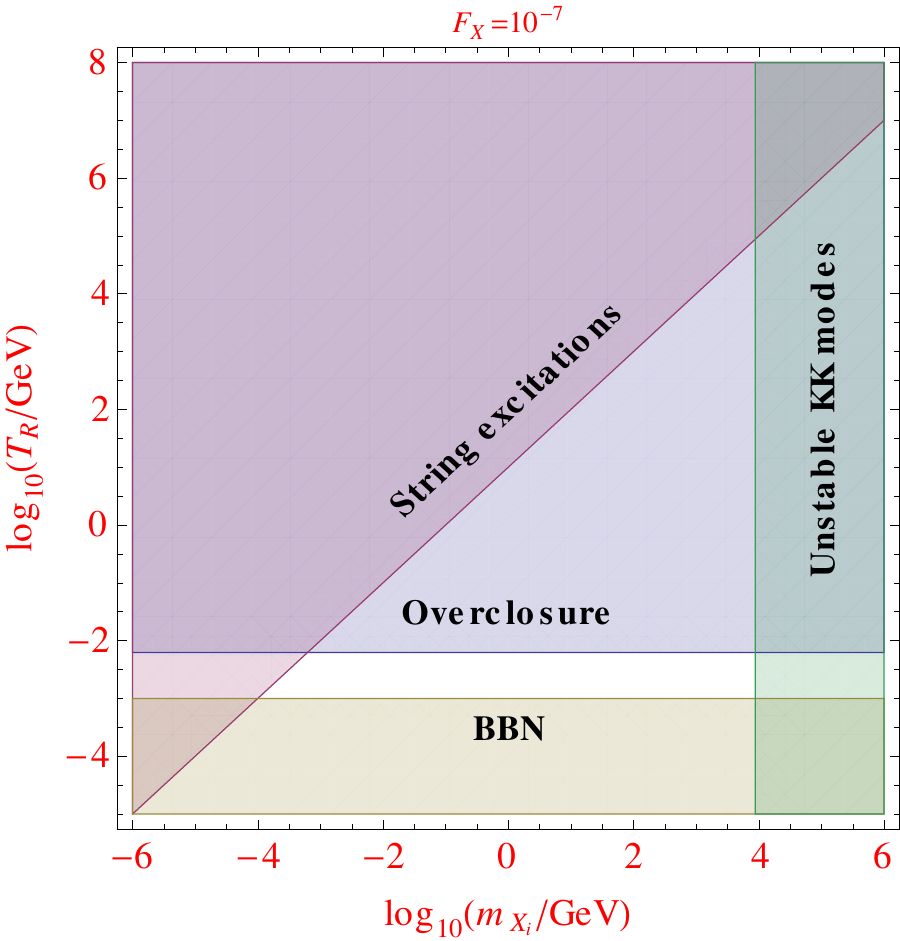}
\hspace{0.1cm}
\includegraphics[width=5.4cm]{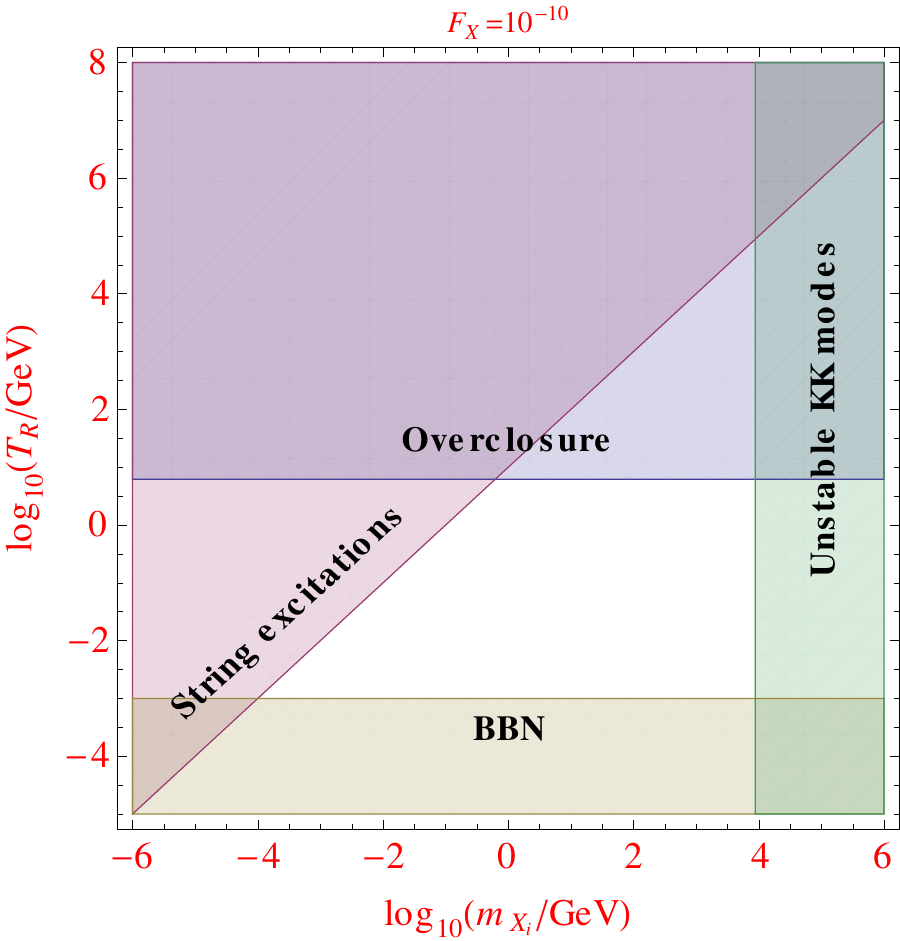}
\hspace{0.1cm}
\includegraphics[width=5.4cm]{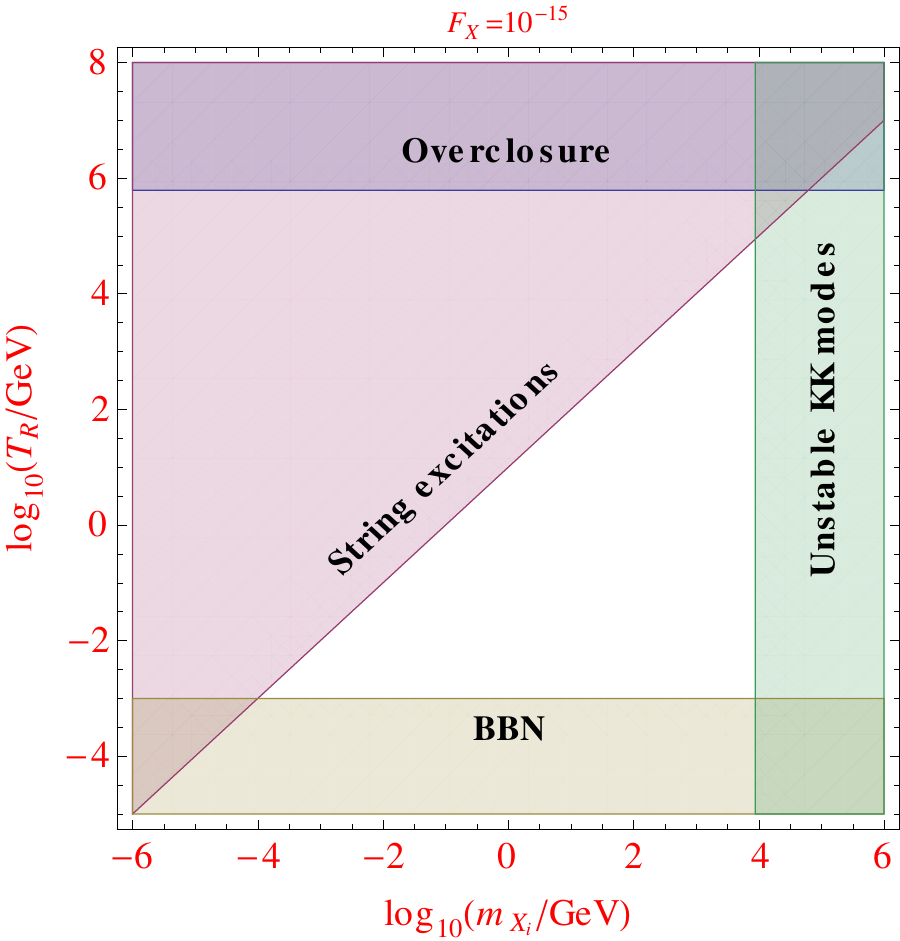}
\caption{Examples of the allowed range, shown as the white (unshaded) region, of
  the reheating temperature ($T_R$) and the lightest KK mass in the
  hidden throats ($m_{X_i}$) in the
  bulk reheating case (section \ref{bulkreheatingsec}) for different energy density fractions $F_X$
  and arbitrary number of throats. We
  plot here the case of $AdS_5 \times X_5$ RS-like throat geometry
  (for other cases see text). Here we have chosen 
  the string scale to be   
  $M_{\rm PL}/100$ and the radius of curvature in each throat to be
  $10M_{\rm PL}^{-1}$, where $M_{\rm PL}=2.4\times 10^{18}$ GeV  
  is the reduced Planck mass.}
\label{fig:4}
\end{figure*}

We first discuss the results for the case when the throat
geometry is well approximated by an $AdS_5 \times X_5$ RS-like
picture. For illustration, this case is shown in Fig.~\ref{fig:4} for the choice
$R_{i} \simeq R_{{\rm in}} = 10 \ell_s$ and $M_s \sim 10^{16}$ GeV.    
Note that in the bulk-reheating case, if there are
long-lived KK modes in the throats (as shown by the dashed-region in Fig.~\ref{fig:4}), 
the allowed mass range for the KK modes 
is even smaller, i.e. between 0.1 MeV-100 GeV. 

Varying $R_i$ and $M_s$, or considering the non-RS-like geometries such
as in~\cite{tye}, leads to the same variation of scales and shrinking
of the allowed region discussed in the previous subsection. However in this case, 
it is not possible to obtain in a model-independent manner a minimum
absolute lower bound on $m_{X_i}$ or a relation between 
the maximum $T_R$ and the scale of inflation.

\subsection{Number of DM candidates and fine-tuning}

Finally, we present an estimate of the total number of the KK dark
matter candidates for the multi-throat scenarios we have
discussed. From 
the top-down approach, given a number $C$ of conifold points in the
Calabi-Yau manifold, the average number 
of throats with a warping $h < \bar h$  is~\cite{march} 
 \beq \label{averageNthroats}
  \bar{N}(h < \bar h|C) \simeq {C \ov 3\log{(1/\bar h)}}~.
 \eeq
Using the upper bound $h_{\rm max}$ on the warping factor for a given
throat to harbor the stable KK modes as derived with the above
analysis, we obtain $\bar{N}(h<h_{\rm max}|C) \sim 0.01 C$, i.e., the 
number of dark matter candidates in our multi-throat scenario is
roughly 1\% of the total number of conifold points in the Calabi-Yau
manifold which depends on the underlying string compactification
model. This also gives a measure of the fine-tuning necessary to
obtain the suitable models of DM we have been considering.

\section{Some Phenomenological Implications}\label{phenoimpsec}

In this section, we briefly discuss some interesting
phenomenological consequences of the multiple DM scenario advocated
here. Our discussion in this section will be qualitative only, since
we cannot present precise quantitative predictions without resorting
to a particular model. For detailed quantitative aspects of the
multiple DM phenomenology, see e.g.,~\cite{otherMDM,DDM,otherpheno}.   
 
(i) The total DM
relic density will be given by the sum of the relic densities in  
the visible and hidden sectors:
 \begin{equation}
  \Omega^{\rm total}_{\text{DM}}h^2=\Omega^{\rm visible}_{\text{DM}}h^2+\Omega^{\rm hidden}_{\text{DM}}h^2\,,
 \end{equation}
where $\Omega^{\rm hidden}_{\text{DM}}h^2$ includes the contribution of the KK 
dark matter particles and, possibly, other stable and weakly interacting
matter from other hidden sectors.
 This implies that the lower limit of the WMAP measured value:
 $\Omega_{\rm CDM} h^2>0.09$~\cite{wmap} is no more applicable to the
 visible sector DM candidate. This releases a large part of the
 parameter space in many visible sector DM models. For example, this
 provides one way to salvage the constrained MSSM with neutralino LSP
 in the light of the recent LHC data, even though the additional
 parameter space available for $\Omega h^2<0.09$~\cite{howie} is not
 spectacularly large. 

 (ii) The KK dark matter candidates in 
our multi-throat scenario are allowed to be light, and hence, could
possibly serve as warm (or even hot) dark matter  
candidates (if their velocity distribution is the opportune one). 
If the KK modes are sufficiently light (of order eV), they
could act as extra sterile degrees of freedom (similarly to sterile
neutrinos~\cite{ben}) from the point of view of the visible sector,
thus possibly explaining the deviation of the measured value of the
effective number of 
relativistic degrees of freedom at BBN, $N_{\rm eff}^{\rm WMAP}=4.34\pm
0.87$~\cite{wmap}, from the SM value $N_{\rm eff}^{\rm SM}=3$. One of
the main constraints on the number of sterile degrees of freedom is provided by the
high sensitivity of the expansion rate 
of the Universe at the BBN temperature: $T_{\rm SM,BBN}\sim$ 1
MeV. Sufficiently light KK modes could change the expansion rate of
the Universe and thereby impact the BBN, even though they do not
reside in the SM throat, and do not have any SM interactions. This
can be translated into a bound (see e.g.~\cite{feng-review}) 
 \beq
  g_{\rm KK}\left(\frac{T_{\rm KK,BBN}}{T_{\rm SM,BBN}}\right)^4 = \frac{7}{8}\cdot 2\cdot (N_{\rm eff}-3)  
 \eeq
where $g_{\rm KK}$ is the number of internal degrees of freedom for the
relativistic KK modes. This bound could be satisfied in our scenario
depending on the precise value of the KK dark matter temperature in the hidden throats.

(iii) In the scenario where inflation occurs in the initial throat, 
the bounds on the local mass scale also constrain the scale of
inflation in the compactified effective theory (for which we must have 
$R_i > \ell_S$, $M_s < M_{\rm PL}$). In particular, when the throat geometry is well
approximated by the $AdS_5 \times X_5$ RS-like picture, the local
string scale, and hence the scale of inflation, in the initial throat is
found to be $\lsim 10^{13}$ GeV. This would put a strong constraint 
on building any inflationary model with large tensor-to-scalar ratio, 
which would generate detectable gravitational waves during or after inflation 
 (see e.g.~\cite{inflation-rev}). Therefore, if such a large tensor-to-scalar ratio
is observed (e.g., by the Planck satellite),  
this would certainly disfavor the multi-throat warped models with RS-like throat
geometry where inflation occurs in a hidden throat, 
since the Universe will be inevitably overclosed by the KK relics. 

(iv) Depending on the model construction, the KK modes residing in
the SM throat could decay to light supergravity modes and/or directly
to the SM particles, thus leading to some interesting indirect
detection signals~\cite{cline,chen}. 
Very light modes in the SM throat
could also provide observable effects for example in the gravity
sector, and hence, a test for our model.

\section{Conclusion} \label{conclusion}

We have discussed the possibility of having multiple dark matter
candidates in the form of the lightest KK-modes residing  
in various long throats which arise naturally out of generic Calabi-Yau
compactification in string theories. The stability of
these KK-modes in long enough throats is due to the small tunneling
rate because of the background warped geometry.
Compliance with cosmological and consistency constraints imposes
strong bounds on the underlying string compactified models.
However, we have shown that there exists some parameter space where
the KK modes can be DM candidates.  We have
been able to study these features in a rather model-independent way,
and thus in the optic of a string landscape analysis.

This gives us a very different
picture for particle dark   
matter than the usually discussed {\it single} WIMP scenario. We also note here 
that relaxing the absolute cosmological stability constraint for the
DM states and instead requiring only  
a dynamical balance against their abundance, as in the so-called Dynamical DM 
scenario~\cite{DDM}, might open up a larger parameter space in our case.

As a word of caution, we must keep in mind the generality of the
models we have considered, by neglecting the precise couplings between
the reheating and the reheated degrees of freedoms in the
bulk-reheating scenario, which preclude us 
from knowing the exact branching ratios, and
assuming pure KK matter content in the
hidden throats and not other possible more complicated matter
sectors as well as radiation.
We leave some of these issues for future investigation.

\begin{acknowledgments}
We thank the referee for a critical reading of the manuscript and for many useful comments. The work of DC is supported by the Belgian National ``Fond de la
Recherche Scientifique'' F.R.S.-F.N.R.S. with a contract ``charg\'e 
de recherche''. The work of BD and AM is supported by the
Lancaster-Manchester-Sheffield 
Consortium for Fundamental Physics under STFC grant ST/J000418/1. 
\end{acknowledgments}



\end{document}